\documentclass[aps,prl,onecolumn]{revtex4-2}
\usepackage[utf8]{inputenc}
\usepackage{bm}
\usepackage{amssymb}
\usepackage{mathtools}
\usepackage{amsmath}
\usepackage{xcolor}
\usepackage{enumitem}
\usepackage{natbib}
\usepackage{hyperref}
\usepackage{lipsum}
\usepackage[ruled,vlined]{algorithm2e}

\begin{document}
\title{Evolution of robustness in growing random networks}
\author{Melvyn Tyloo}
\affiliation{Theoretical Division and Center for Nonlinear Studies (CNLS), Los Alamos National Laboratory, Los Alamos, NM 87545, USA}
\date{\today}
\begin{abstract}
    Networks are widely used to model the interaction between individual dynamical systems. In many instances, the total number of units as well as the interaction coupling are not fixed in time, but rather constantly evolve. In terms of networks, this means that the number of nodes and edges change in time. Various properties of coupled dynamical systems essentially depend on the structure of the interaction network, such as their robustness to noise. It is therefore of interest to predict how these properties are affected when the network grows and what is their relation to the growing mechanism. Here, we focus on the time-evolution of the network's Kirchhoff index. We derive closed form expressions for its variation in various scenarios including both the addition of edges and nodes. For the latter case, we investigate the evolution where a single node with one and two edges connecting to existing nodes are added recursively to a network. In both cases we derive relations between the properties of the nodes to which the new one connects, and the global evolution of the network robustness. In particular, we show how different scalings of the Kirchhoff index as a function of the number of nodes are obtained. We illustrate and confirm the theory with numerical simulations of growing random networks.
\end{abstract}

\maketitle


\setcounter{section}{0} 

\section{Introduction}
Complex networks are broadly used to model interaction within natural and engineered systems~\cite{strogatz2001exploring,pikovsky2002synchronization,newman2018networks}. They describe the interaction taking place between individual elements, such as the chemical bonds between atoms that form a molecule, or the communication transmitted between neighboring individuals in flocks of birds or vehicular platoons~\cite{ARENAS200893}. From their structure important properties of the coupled dynamical systems can be deduced such as their intrinsic natural frequencies or their stability and robustness to external perturbations~\cite{BOCCALETTI2006175}. While in many instances the structure of the coupling network as well as the number of interacting elements composing the system remain constant in time, it is typically not the case in a wide variety of coupled systems such as social networks, vehicular platoon formation, swarming autonomous robots, animal collective behaviors, cells evolution, molecules interacting in chemical reactions~\cite{dorogovtsev2003evolution,newman2001clustering,zhao2011entropy,Porter2016,della2022synchronization}. In all these examples, when an element (commonly represented as a node) or an interaction (represented as an edge) is added to or removed from the system, its overall dynamical properties are modified. In particular, both the steady states and the corresponding transient stability are affected by the evolution of the system. It is therefore an important task to predict how these properties are changing while the network evolves, and be able to anticipate potential development of instabilities. More specifically, if one has to sequentially add agents to a system, how should they interact with the existing units so that stability is preserved, or at least not hindered to importantly. This is the main question that we investigate in this manuscript. 
Previous works have considered the evolution of some network properties such as the degree distribution in random growing networks with preferential attachment ~\cite{krapivsky2000connectivity,dorogovtsev2000structure}, or the evolution of the Wienner index in random recursive trees~\cite{neininger2002wiener}. In this manuscript, we investigate the time-evolution of the \textit{Kirchhoff index}~\cite{klein1993resistance,bonchev1994molecular,https://doi.org/10.1002/qua.10057}. The latter has proved to be useful in chemistry~\cite{doi:10.1246/bcsj.44.2332,klein1993resistance,mohar1993novel} and networked dynamical systems~\cite{tyloo2018robustness,baumann2020laplacian}. For coupling networks that are not growing and which are static in time, the robustness of diffusively coupled oscillators have been direclty related to the Kirchhoff index of the effective coupling network~\cite{tyloo2018robustness,tyloo2019key,ronellenfitsch2018optimal}. Namely, the larger the Kirchhoff index, the more important are the fluctuations within the dynamical system. Indeed, let us consider a set of $N$ oscillators each with a continuous degree of freedom $x_i\in\mathbb{R}$\,, diffusively coupled together and subjected to noise as,
\begin{eqnarray}
\dot{x}_i = -\sum_{j=1}^N a_{ij}(x_i - x_j) + \eta_i\,, \quad i=1,...N\,,
\end{eqnarray}
where $a_{ij}=a_{ji}>0$ are the elements of the adjacency matrix encoding the undirected coupling network, $\eta_i$'s are white uncorrelated noise inputs, i.e. $\langle \eta_i(t)\eta_j(t') \rangle = \eta_0^2\,\delta_{ij}\,\delta(t-t)$\,. Then, the average variance in the long time limit is given by~\cite{Tyloo_2022},
\begin{eqnarray}\label{eqosc}
\frac{1}{N}\sum_{j=1}^N\langle x_j^2 \rangle = \frac{\eta_0^2}{2} Kf_1/N\,,
\end{eqnarray}
with $Kf_1$ the Kirchhoff index of the coupling network (see Sec.~\ref{sec2} for the definition). Similar relations are obtained for deterministic perturbations that have a short correlation time~\cite{tyloo2018robustness}. Given this direct connection between a global network index and the fluctuations of the dynamical system supported by the network, it is interesting to investigate how the Kirchhoff index evolve while the network grows. For the evolution of the network, we consider a simple growing algorithm where at each iteration, a single new node is added and connects to existing nodes. We derive analytical expression for the time-evolution of the Kirchhoff. In particular, when connecting the new node to the existing ones, we identify which of their nodal properties influence the scaling of the Kirchhoff index as a function of the total number of nodes. One can then use these properties when adding new nodes to achieve different scalings for the Kirchhoff index and thus, for the fluctuations.\\
The manuscript is organized as follows. In Sec.~\ref{sec2}, we give the definition of the Kirchhoff index and discuss bounds previously derived. In Sec.~\ref{sec3}, we consider growing networks and provide expressions for the time-evolution of the Kirchhoff index when edges and nodes are added. Finally, Sec.~\ref{sec4} gives our conclusions and outlook.

\section{Kirchhoff index}\label{sec2}
\subsection{Definitions}
Let us consider a graph (called network in the following) $G$ made of $N$ vertices (called nodes in the following) and $M$ edges. Each edge $\epsilon_{(ij)}$ between two nodes $i$ and $j$ has an associated weight $a_{ij}>0$\,. The network Laplacian matrix is commonly defined as $L$ as $L_{ij}=-a_{ij}$ if $i\neq j$ and there exists an edge between nodes $i$ and $j$\,, and $L_{ij}=0$ otherwise, and $L_{ii}=\sum_{k=1}^Na_{ik}$ for $i=1,...N$\,.
The Kirchhoff index ($Kf_1$) of an undirected network is given by the sum of the effective resistance distances ($\Omega_{ij}$) between all the nodes, i.e.\cite{klein1993resistance}
\begin{equation}
    Kf_1=\sum_{i<j}\Omega_{ij}\,.
\end{equation}
The resistance distance between node $i$ and $j$ is defined by,
\begin{equation}
    \Omega_{ij} = [\mathbb{L}^\dagger]_{ii} -2[\mathbb{L}^\dagger]_{ij} + [\mathbb{L}^\dagger]_{jj}\,,
\end{equation}
where $\mathbb{L}^\dagger$ denotes the pseudo-inverse of the Laplacian matrix $\mathbb{L}$ of the network. Using the eigenvectors ${\bf u}_{\alpha}$ and eigenvalues $\lambda_1=0<\lambda_2<...<\lambda_N$\,, of $\mathbb{L}$\,, one conveniently rewrite the resistance distance $\Omega_{ij}=\sum_{\alpha>1}(u_{\alpha,i}-u_{\alpha,j})^2/\lambda_\alpha$\,, which yields for the Kirchhoff index~\cite{lukovits1999resistance},
\begin{equation}\label{Kf}
    Kf_1=N\sum_{\alpha>1}\frac{1}{\lambda_\alpha} = N\,{\rm Tr}[\mathbb{L}^\dagger]\,.
\end{equation}
Depending on the time-scale of the noise input, the amplitude of the small fluctuations of diffusively coupled oscillators can be expressed in terms of the Kirchhoff index or its generalization which reads~\cite{tyloo2018robustness},
\begin{equation}
    Kf_p=N\sum_{\alpha>1}\frac{1}{\lambda_\alpha^p} =  N\,{\rm Tr}[{\mathbb{L}^\dagger}^p]\,.
\end{equation}
For specific network models for which the spectrum is known, the Kirchhoff can be analytically obtained. For example, for the complete, star and cycle network one has $Kf_1 \cong N, N^2, N^3$\,, and $Kf_2\cong 1, N^2, N^5$ respectively as the number of nodes $N$ becomes large. These network models will be useful below when we consider the limiting case of random growing networks. In the specific case where the network is a tree, the resistance distance is equal to the shortest path distance in the same network where all the weights on the edges have been replaced by their inverse. In such a situation, the Kirchhoff index is equal to the Wienner index~\cite{doi:10.1246/bcsj.44.2332}, that is defined as the sum of all shortest path distances in the network. In the following we discuss the Kirchhoff index, as the growing networks we consider do not have to be trees.
From the resistance distance, one also defines a centrality measure that reads,
\begin{eqnarray}\label{cendef}
    C(i) = \left[ \sum_{j=1}^N\Omega_{ij}/N \right]^{-1} = \left[\mathbb{L}^\dagger_{ii}+Kf_1/N^2\right]^{-1}\,,
\end{eqnarray}
where $C(i)$ is called the resistance centrality of node $i$\,.\\

\subsection{Lower bound on $Kf_1$}
The Kirchhoff index has been extensively studied and many bounds depending on the number of nodes $N$\,, edges $n_e$ and other properties have been derived. Relevant for the following, is a lower bound obtained by Zhou and Trinajsti\'{c}~\cite{ZHOU2008120} which is: for a connected network with $N\ge3$\,, $n_e$ edges and a maximum degree $\Delta$\,, then 
\begin{eqnarray}\label{ineq}
    Kf_1(N)\ge \frac{N}{1+\Delta} + \frac{N(N-2)^2}{2N_e -1 -\Delta}\,.
\end{eqnarray}
From this inequality, one concludes that, as long as $n_e\propto N$ , then $Kf_1/N$ scales at least as $N$ when the number of nodes becomes large. This is the case in the growing algorithm we investigate below, where a single new node is added at each iteration, such that $n_e\propto N$\,. Therefore, the lowest scaling achievable for $Kf_1/N$ within our growing algorithm is linear in $N$\,.

\section{Robustness of growing networks}\label{sec3}
Networks can grow in two ways: (i) some new nodes are connected to the existing network nodes; (ii) some edges are added within the existing nodes. On one hand, for (i) intuitively, based on the examples given in Sec.~\ref{sec2} and Eq.~(\ref{ineq}), the Kirchhoff index increases at least linearly with $N$\,. On the other hand, for (ii) one can show that the Kirchhoff index can only decrease by adding a new edge in the network. Indeed, adding one edge with corresponding weight $a_{kl}>0$ between nodes $k$ and $l$ is a rank-1 modification of the Laplacian matrix, i.e. $\mathbb{L}{(t+1)}=\mathbb{L}{(t)} + a_{kl}{\bm e}_{kl}{\bm e}_{kl}^\top$\,, where $[e_{kl}]_{i} = (\delta_{ik} - \delta_{il})\in\mathbb{R}^{N_t}$\,, with $N_t$ the number of nodes at iteration $t$\,. Therefore, if the Kirchhoff index at iteration $t$ is $Kf{(t)}$\,, one can use the Sherman-Morrison-Woodbury formula~\cite{golub2013matrix,doi:10.1137/1.9781611975031.153} to obtain the Kirchhoff index at step $t+1$\,,
\begin{eqnarray}\label{smw}
    Kf_k{(t+1)} = Kf_1{(t)} - \frac{a_{kl}{\rm Tr}[\mathbb{L}^\dagger {\bm e}_{kl}{\bm e}_{kl}^\top \mathbb{L}^\dagger]}{1+a_{kl}\Omega_{lk}{(t)}}\, = Kf_1{(t)} - \frac{a_{kl}\Omega_{kl}^{(2)}(t)}{1+a_{kl}\Omega_{kl}{(t)}}\,,
\end{eqnarray}
where $\Omega_{kl}^{(2)}(t) = \sum_{\alpha>1}(u_{\alpha,i}-u_{\alpha,j})^2/\lambda_\alpha^2$ is a semi-metric~\cite{tyloo2019key}.
As both $\Omega_{kl}^{(2)}(t)$ and $\Omega_{kl}{(t)}$ are always positive, $Kf_1$ can only decrease when an edge is added to the existing network. 
Below, we discuss how the Kirchhoff index is modified when a single node is added to the existing network together with $m$ new edges.

\begin{figure}
\centering
\includegraphics[scale=0.9]{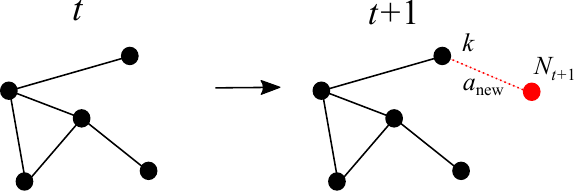}
\caption{Evolution of the network from iteration $t$ to $t+1$\,, where a new node (in red) connecting to a single existing one (in black), $k$\, has been added. The label of the new node is $N_{t+1}=N_t + 1$\,. No new path is create within the existing nodes.}\label{fig1}
\end{figure}
\subsection{One new node with a single connection ($m=1$)}\label{sec_s1}
\begin{figure}
\centering
\includegraphics[scale=0.75]{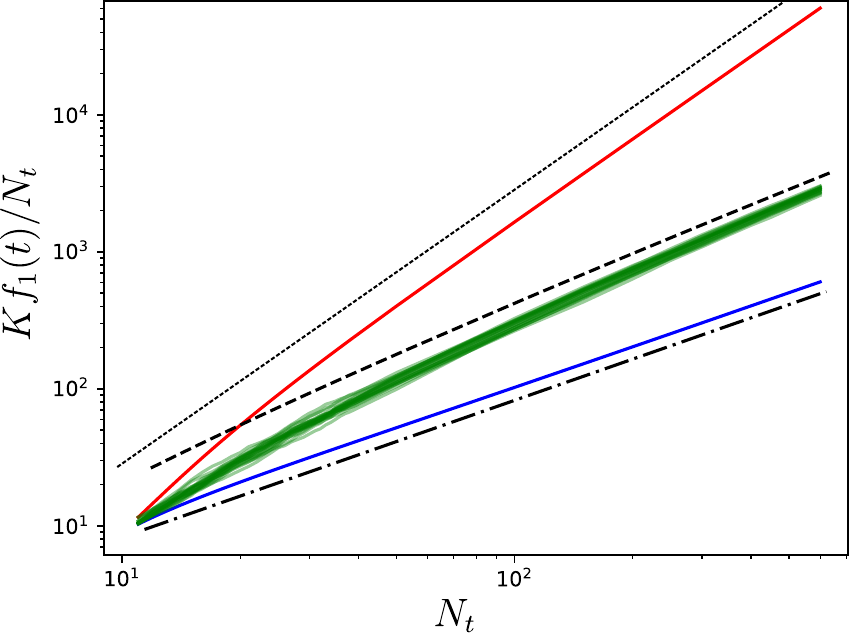}
\caption{Evolution of the Kirchhoff index divided by the number of nodes $N_t$ when a new node connects to a single existing one at each iteration. The initial network has 10 nodes and is obtained from a Watts-Strogatz rewiring procedure with nearest neighbors coupling~\cite{watts1998collective}. The green curves correspond to 20 realizations where one starts from the initial network, to which nodes are then recursively added, uniformly selecting at random the existing nodes to which they connect. For large $N_t$\,, the green curves follow the scaling of Eq.~(\ref{eqtree}). The red and blue curves are obtained by selecting at each iteration the least and most central existing node, respectively. When $N_t$ is large, they follow the scalings of Eqs.~(\ref{m1p}), (\ref{m2p})\,. The dotted, dashed, dotted-dashed black lines give $N_t^2$\,, $N_t\,{\log}N_t$\,, $N_t$\,, respectively.}\label{fig2}
\end{figure}

We investigate the evolution of the Kirchhoff index for the growing process depicted on Fig.~\ref{fig1}\,. When the new node connects to a single existing one, if one starts with a network that is a tree, then the network will remain a tree as it grows. In addition to that, if the selected existing node is uniformly chosen at random, the latter is also called a \textit{random recursive tree}. In such a situation, one can replace the resistance distance by the geodesic or shortest path distance (and thus the Kirchhoff index by the Wienner index) in the following discussion~\cite{neininger2002wiener,WAGNER20121502}. But in general, we do not assume that the starting network is a tree. At iteration $t$ one has the Kirchhoff index $Kf_1{(t)} = \frac{1}{2}\sum_{i,j=1}^{N_t}\Omega_{ij}{(t)}$\,. If the new node at the iteration $t+1$ is connected to node $k$\,, one has,
\begin{eqnarray}\label{m1}
    Kf_1{(t+1)} &=& Kf_1{(t)} + \sum_{l=1}^{N_t}\Omega_{kl}{(t)} +  \frac{N_t}{a_{\rm new}}\,,
\end{eqnarray}
where $a_{k(N_{t+1})}=a_{\rm new}$ is the weight of the newly added edge between nodes $k$ and $N_{t+1}$\,. In this simple case, one observes that the modification of the Kirchhoff index is given by the sum of the resistance distances from node $k$ to all the other already existing nodes in the network plus $N_t$ times the resistance of the newly added edge (see Fig.~\ref{fig1}). The less central node $k$ is (in terms of resistance distances with respect to the existing nodes), the more the Kirchhoff index grows. As expected, $Kf_1{(t)}$ only increases with the iterations, as no new path within the existing nodes is created. If the node to which the new node connects is uniformly selected at random amongst the existing one at each iteration, then on average the Kirchhoff index will increase as,
\begin{eqnarray}\label{m12}
    \langle Kf_1{(t+1)} \rangle &=& \langle Kf_1{(t)}\rangle\left(1 + \frac{2}{N_t}\right) +  \frac{N_t}{a_{\rm new}}\\
&=&  \frac{(N_0+t+1)}{a_{\rm new}}\left[ \frac{a_{\rm new}\frac{Kf_1(0)(N_0+2)}{N_0}(N_0+t+2) -2(N_0+2)(t+1) }{(N_0+1)(N_0+2)} \right.\nonumber\\
&+&\left. (N_0 +t+2)(H_{N_0+t+1} - H_{N_0}) \right]\\
&=& \frac{(N_0+t+1)}{a_{\rm new}}\left[ \frac{a_{\rm new}\frac{Kf_1(0)(N_0+2)}{N_0}(N_0+t+2) -2(N_0+2)(t+1) }{(N_0+1)(N_0+2)} \right.\nonumber\\
&+&\left. (N_0 +t+2)\{\psi_0(N_0+t+1) - \psi_0(N_0)\} \right]\,,\label{eqrec}
\end{eqnarray}
where $N_0$ and $Kf_1(0)$ are respectively, the initial number of nodes and  the initial Kirchhoff index, $H_N=\sum_{k=1}^N k^{-1}$ is the $N$-th harmonic number which can be written as $H_N = \gamma + \psi_0(N+1)$ where $\gamma\cong0.577$ is the Euler-Mascheroni number and $\psi_0(n)=\Gamma'(n)/\Gamma(n)$ is the digamma function. As its integer argument becomes large, the digamma function satifies $\psi_0(n)\overset{n\rightarrow\infty}{\propto}\ln n$\,. Therefore, when the number of iterations becomes large, the last term in Eq.~(\ref{eqrec}) will dominate such that,
\begin{eqnarray}
    \langle Kf_1{(t)} \rangle &\overset{t\rightarrow\infty}{ \propto} N_t^2\,\log N_t\,.\label{eqtree}
\end{eqnarray}
The scaling is confirmed numerically in Fig.~\ref{fig2}, where the solid green curves are 20 realizations of the random growing process where one starts from ten connected nodes and then one new node is added at each iteration, that connects uniformly at random to an existing one. One observes that the simulations follow the predicted scaling of Eq.~(\ref{eqtree}) given by the dashed black line. Note that it is the same scaling as the Wienner index for random recursive trees~\cite{neininger2002wiener}. This random evolution of the network is bounded by the two limiting cases that we now discuss.
Instead of uniformly picking within the existing nodes, one may use some property of the nodes. Let us discuss what happens when one selects the most or least central node in terms of resistance distance. If the least central node $k$ at each iteration $t$\,, i.e. with largest $\sum_{l=1}^{N_t}\Omega_{kl}{(t)}$\,, is chosen to be connected to the new node, the network will tend to form a chain. Therefore, assuming that the weights on the edges are of order $1$\,, when $N_t$ becomes large, one has 
\begin{eqnarray}
    \sum_{l=1}^{N_t}\Omega_{kl}{(t)} \cong \frac{N_t(N_t-1)}{2}\,.
\end{eqnarray}
In this case, the Kirchhoff index grows as,
\begin{eqnarray}\label{m1p}
    Kf_1{(t+1)} &\cong& Kf_1{(t)} + \frac{N_t(N_t-1)}{2}  +  \frac{N_t}{a_{\rm new}}\overset{t\rightarrow \infty}{\propto} N_t^3\,,
\end{eqnarray}
which is faster than in the random uniform case Eq.~(\ref{eqtree})\,. If instead, one selects the most central at each time step, then the network becomes star-like. Indeed, by connecting a new node to the most central existing one, its centrality becomes even more important. This means that all the newly added nodes will connect to the same node. Thus, assuming that the edges weights are of order $1$\,, one has for large $N_t$\,,
\begin{eqnarray}
    \sum_{l=1}^{N_t}\Omega_{kl}{(t)} &\cong& {(N_t-1)}\,,\\
    Kf_1{(t+1)} &\cong& Kf_1{(t)} + (N_t-1)  +  \frac{N_t}{a_{\rm new}}\overset{t\rightarrow \infty}{\propto} N_t^2\,.\label{m2p}
\end{eqnarray}
Interestingly, by selecting the most central node, one achieves a scaling for $Kf_1$ with $N_t$ only $\log N_t$ better than Eq.(\ref{eqtree}) where the node is uniformly chosen amongst the existing ones. 
\subsection*{Discussion}
According to Eqs.~(\ref{eqtree}), (\ref{m1p}), (\ref{m2p}), the weakest growth in the Kirchhoff index is obtained when the new nodes simply connect to the most central existing one in terms of resistance distance. Using this mechanism to grow a network leads to a very specific structure where a single node is connected to almost all the other ones. While such structure enhance the transient stability of the system by minimizing the growth of the small fluctuations, it also makes the system very vulnerable to any failure of this particular most central node. Indeed, if one removes that node from the system, then most of its components will become disconnected. Given this structural weakness, selecting nodes at random when adding new nodes seems like a more robust option, as the growth of the Kirchhoff index is only $\log N_t$ worth than selecting the most central one. Moreover, the connection within the network are more uniformly distributed in that situation, reducing the number of disconnected components in case of failures. On the opposite side, if one wants to increase as much as possible the fluctuations in the system, new nodes should be connected to the least central node in terms of resistance distance.

\begin{figure}
\centering
\includegraphics[scale=0.9]{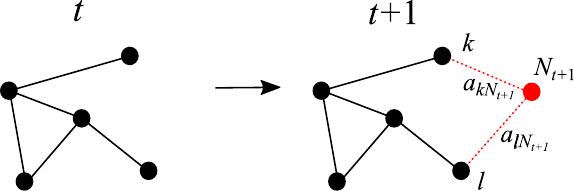}
\caption{Evolution of the network from iteration $t$ to $t+1$\,, where a new node (in red) connecting to two existing ones (in black), $k$ and $l$\,, has been added. The label of the new node is $N_{t+1}=N_t + 1$\,. In this case, a new path between $k$ and $l$ has been created.}\label{fig002}
\end{figure}
\subsection{One new node with two connections ($m=2$)}
\begin{figure}
\centering
\includegraphics[scale=0.77]{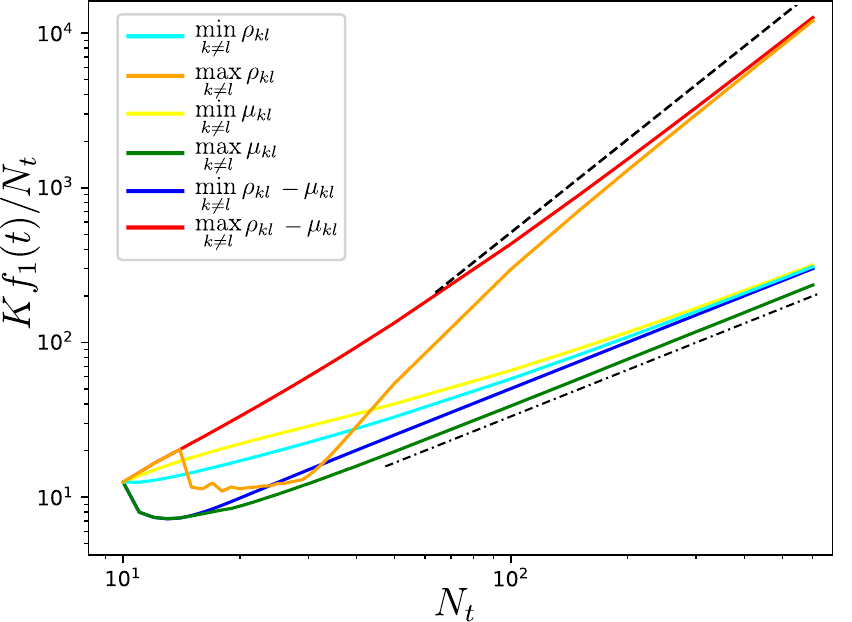}
\caption{Evolution of the Kirchhoff index divided by the number of nodes $N_t$ when a new node connects to two existing ones ($k$ and $l$) at each iteration. Two nodes are selected by minimizing/maximizing $\mu_{kl}(t)$\,, $\rho_{kl}(t)$\,, $\rho_{kl}(t) - \mu_{kl}(t)$ at each iteration. The meaning of each curve is given in the legend in insets. The initial network has 10 nodes and is obtained from a Watts-Strogatz rewiring procedure with nearest neighbors coupling~\cite{watts1998collective}. The black dashed-dotted and dashed lines give the scalings $N_t$ and $N_t^2$\,, respectively. Note that in these simulations we make sure that $k\neq l$\, but found similar scalings when relaxing this condition.}\label{fig4}
\end{figure}

The case where one node is added together with two edges at each iteration is more complex as an increasing number of loops is introduced into the network. If the new node is connected to the existing nodes $k$ and $l$\,, the effective resistance along the new path going from $k$ to $l$ is,
\begin{eqnarray}
    \omega_{kl}=a_{k\,N_{t+1}}^{-1} + a_{l\,N_{t+1}}^{-1}\,.
\end{eqnarray}
The process is depicted on Fig.~\ref{fig002}.
It is important to remark that, adding node $N_{t+1}$ is not the same as adding an edge between nodes $k$ and $l$ whose weight would be $\omega_{kl}^{-1}$\,. However, replacing the path where the new node is located by an equivalent edge provides a lower bound on the new Kirchhoff index. Indeed, by doing so, one obtains the sum of new resistance distances between the already existing nodes,
\begin{eqnarray}\label{eqm21}
    Kf_k{(t+1)} &=& \frac{1}{2}\sum_{i,j=1}^{N_t}\Omega_{ij}{(t+1)} + \sum_{i=1}^{N_t}\Omega_{iN_{t+1}}{(t+1)}\\
 &=&    Kf_1{(t)} - \frac{\omega_{kl}^{-1}\,\Omega_{kl}^{(2)}(t)}{1+\omega_{kl}^{-1}\Omega_{lk}{(t)}} + \sum_{i=1}^{N_t}\Omega_{iN_{t+1}}{(t+1)}\,.\label{eqm2}
\end{eqnarray}
The variation of the Kirchhoff index is a function of $\Omega_{kl}^{(2)}(t)$ and of how central in terms of resistance distance the new node is [see last term in Eq.~(\ref{eqm2})]. For the latter term, it is challenging to find a closed form expression. However, one can have an estimate of it based on the resistance distance in the new network. Indeed, once the new node has been added, the resistance distances within the existing nodes at iteration $t$ become
\begin{eqnarray}\label{res2}
    \Omega_{ij}(t+1)= \Omega_{ij}{(t)} - \frac{\omega_{kl}^{-1}\,[{\bm e}_{ij}^\top\mathbb{L}^\dagger(t) {\bm e}_{kl}]^2}{1+\omega_{kl}^{-1}\Omega_{kl}{(t)}}\,,\quad i,j=1,...,N_t\,,
\end{eqnarray}
where we replaced the new node by an equivalent edge between $k$ and $l$ using the Sherman-Morrison-Woodbury formula similarly as in Eq.~(\ref{smw})\,. Using Eq.~(\ref{res2}), one can approximate the last term in Eq.~(\ref{eqm2}) as the weighted average,
\begin{eqnarray}\label{eqrtpu}
 \sum_{i=1}^{N_t}\Omega_{iN_{t+1}}{(t+1)} &\cong& \frac{1}{(a_{k\,N_{t+1}} + a_{l\,N_{t+1}})}\sum_{j=1}^{N_t}\left[a_{kj}\Omega_{kj}{(t+1)} + a_{lj}\Omega_{lj}{(t+1)} \right]\,.
\end{eqnarray}
We expect this approximation to be valid when the edge weights surrounding the new node, including $a_{k\,N_{t+1}}$\,, $a_{l\,N_{t+1}}$ are homogeneous enough, or when $a_{k\,N_{t+1}}$\,, $a_{l\,N_{t+1}}$ are much larger than the surrounding edge weights. Indeed, if the $a_{k\,N_{t+1}}$\,, $a_{l\,N_{t+1}}$ are weak, the centrality of the new node is expected to be lower than those of $k$ and $l$\,.
Using this approximation together with Eq.~(\ref{eqrtpu}), yields for Eq.~(\ref{eqm2}),
\begin{eqnarray}\label{eqm13}
  Kf_1{(t+1)} &\cong&  Kf_1{(t)}\\
  -\frac{\omega_{kl}^{-1}}{1+\omega_{kl}^{-1}\Omega_{kl}{(t)}} && \hspace{-0.4cm}\left\{ \Omega_{kl}^{(2)}(t) + \frac{\sum_{j=1}^{N_t}\left[a_{k\,N_{t+1}}({\bm e}_{kj}^\top\mathbb{L}^\dagger(t) {\bm e}_{kl})^2 + a_{l\,N_{t+1}}({\bm e}_{lj}^\top\mathbb{L}^\dagger(t) {\bm e}_{kl})^2\right]}{(a_{k\,N_{t+1}} + a_{l\,N_{t+1}})} \right\}\nonumber  \\
  &+& \frac{1}{(a_{k\,N_{t+1}} + a_{l\,N_{t+1}})}\sum_{j=1}^{N_t}\left[a_{k\,N_{t+1}}\Omega_{kj}{(t)} + a_{l\,N_{t+1}}\Omega_{lj}{(t)}\right]\nonumber\\
&=& Kf_1{(t)}-\frac{\omega_{kl}^{-1}}{1+\omega_{kl}^{-1}\Omega_{lk}{(t)}}  \,\left\{ 2\Omega_{kl}^{(2)}(t) \right.\\
&+& N_t\left.\frac{a_{k\,N_{t+1}}\left[\mathbb{L}^\dagger_{kk}(t)-\mathbb{L}^\dagger_{kl}(t)\right]^2+ a_{l\,N_{t+1}}\left[\mathbb{L}^\dagger_{ll}(t)-\mathbb{L}^\dagger_{lk}(t)\right]^2}{(a_{k\,N_{t+1}} + a_{l\,N_{t+1}})} \right\}\nonumber  \\
   &+& N_t \frac{a_{k\,N_{t+1}}C^{-1}{(k,t)} + a_{l\,N_{t+1}}C^{-1}{(l,t)}}{(a_{k\,N_{t+1}} + a_{l\,N_{t+1}})}\nonumber\\
 &=& Kf_1{(t)}-\frac{\omega_{kl}^{-1}}{1+\omega_{kl}^{-1}\Omega_{lk}{(t)}}  \,\left\{ 2\Omega_{kl}^{(2)}(t) \right.\\
 && \hspace{-2cm}+N_t\left.\frac{a_{k\,N_{t+1}}\left[\Omega_{kl}(t)  + C^{-1}(k,t) - C^{-1}(l,t)\right]^2+ a_{l\,N_{t+1}}\left[\Omega_{kl}(t) + C^{-1}(l,t) - C^{-1}(k,t)\right]^2}{2(a_{k\,N_{t+1}} + a_{l\,N_{t+1}})} \right\}\nonumber  \\
  &+& N_t\frac{a_{k\,N_{t+1}}C^{-1}{(k,t)} + a_{l\,N_{t+1}}C^{-1}{(l,t)}}{(a_{k\,N_{t+1}} + a_{l\,N_{t+1}})}\nonumber\\ 
  &=& Kf_1{(t)}-\frac{\omega_{kl}^{-1}}{1+\omega_{kl}^{-1}\Omega_{lk}{(t)}}  \,\left\{ 2\Omega_{kl}^{(2)}(t) + N_t\frac{\Omega^2_{kl}(t)}{2}  \right.\nonumber\\
 &+& N_t\frac{\left[ C^{-1}(k,t) - C^{-1}(l,t) \right]^2}{2} \nonumber  \\
  &+& \left. N_t\frac{(a_{k\,N_{t+1}}- a_{l\,N_{t+1}})}{(a_{k\,N_{t+1}} + a_{l\,N_{t+1}})}{\left[ C^{-1}(k,t) - C^{-1}(l,t) \right]}\Omega_{kl}(t)  \right\} \nonumber\\
  &+& N_t\frac{a_{k\,N_{t+1}}C^{-1}{(k,t)} + a_{l\,N_{t+1}}C^{-1}{(l,t)}}{(a_{k\,N_{t+1}} + a_{l\,N_{t+1}})}\label{eqIM}
\end{eqnarray}
where we used the relation between $\mathbb{L}^\dagger_{ii}$ and the resistance centrality of node $i$ [see Eq.~(\ref{cendef})].
Now this expression gives an approximation of $Kf_1(t+1)$ based only on quantities at iteration $t$\,.  To reduce the increase of $Kf_1$\,, one should therefore find nodes $k$ and $l$ such that $\Omega_{kl}^{(2)}(t)$ and $\Omega_{kl}(t)$ are large, and which have very different resistance centralities, e.g. $k$ being part of the most central nodes while $l$ belongs to the least central ones.
We group together the terms in Eqs.~(\ref{eqIM}) as,
\begin{eqnarray}
     \mu_{kl}(t)&=&\frac{\omega_{kl}^{-1}}{1+\omega_{kl}^{-1}\Omega_{lk}{(t)}}\left\{ 2\Omega_{kl}^{(2)}(t) + N_t\frac{\Omega^2_{kl}(t)}{2} + N_t\frac{\left[ C^{-1}(k,t) - C^{-1}(l,t) \right]^2}{2} \right.\nonumber\\
 &+&\left.N_t\frac{(a_{k\,N_{t+1}}- a_{l\,N_{t+1}})}{(a_{k\,N_{t+1}} + a_{l\,N_{t+1}})}{\left[ C^{-1}(k,t) - C^{-1}(l,t) \right]} \Omega_{kl}(t) \right\}\,,\label{eqmu}\\
 \rho_{kl}(t) &=&N_t\frac{a_{k\,N_{t+1}}C^{-1}{(k,t)} + a_{l\,N_{t+1}}C^{-1}{(l,t)}}{(a_{k\,N_{t+1}} + a_{l\,N_{t+1}})}\,. \label{eqrho}
\end{eqnarray}
Then, one can choose nodes $k$ and $l$ that minimize/maximize the latter quantities. To get more intuition, we investigate numerically Eqs.~(\ref{eqmu}) and (\ref{eqrho}). In particular, we consider the maximization or minimization at each iteration of $\mu_{kl}(t)$\,, $\rho_{kl}(t)$\,, $\rho_{kl}(t) - \mu_{kl}(t)$\,.
This is shown in Fig.~\ref{fig4}. We consider edge weights such that $a_{k\,N_{t+1}}=a_{l\,N_{t+1}}=1$\,, meaning that the last term in $\mu_{kl}(t)$ vanishes. As expected, maximizing $\mu_{kl}(t)+\rho_{kl}(t)$ (red curve) at each iteration gives the most important increase in $Kf_1(t)/N_t$ which scales as $N_t^2$\,. The same scaling is obtained if one maximizes only $\rho_{kl}(t)$ (orange curve). The minimization of $\mu_{kl}$ (yellow curve) does not produce such an increase of the Kirchhoff index, which seems to remain linear i.e. $Kf_1(t)/N_t \propto N_t$ as $t$ becomes large. A similar linear scaling is observed for the minimization of $\rho_{kl}(t) - \mu_{kl}(t)$ (blue curve), $\rho_{kl(t)}$ (cyan curve) and the maximization of $\mu_{kl}(t)$ (green curve). Interestingly, one observes that the maximization of $\mu_{kl}(t)$ gives a lower Kirchhoff index than the minimization of $\rho_{kl}(t) - \mu_{kl}(t)$\,. Therefore, one can tune the increase of the Kirchhoff index by choosing one or another quantity to optimize at each iteration.
\begin{figure}
\centering
\includegraphics[scale=0.77]{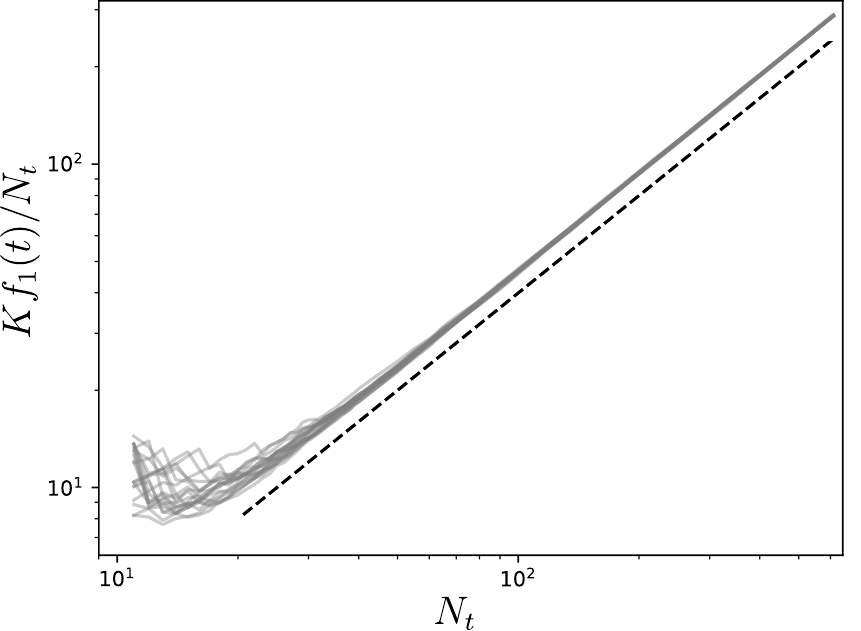}
\caption{Evolution of the Kirchhoff index divided by the number of nodes $N_t$ when a new node connects to two existing ones at each iteration. The two nodes are selected uniformly at random amongst the existing ones at each iteration. Each grey line (20 in total) is one realization of the process. The initial network has 10 nodes and is obtained from a Watts-Strogatz rewiring procedure with nearest neighbors coupling~\cite{watts1998collective}. The black dashed line gives the linear scaling with $N_t$\,.}\label{fig5}
\end{figure}
In Fig.~\ref{fig5}, we simulate the case where $k$ and $l$ at uniformly chosen are random at each iteration. One observes that the 20 realizations of the process yield a linear scaling of $Kf_1(t)/N_t$ with $N_t$\,.

\subsection*{Discussion}
In Sec.~\ref{sec_s1}, we saw that selecting uniformly at random the existing node to which the new one connects, is $\log N_t$ worth for the growth of the Kirchhoff index compared to selecting the most central one. Quite interestingly here, when the new nodes connect to two existing ones, selecting them uniformly at random produces that same scaling as the one obtained by minimizing the relevant quantity $\rho_{kl}(t) - \mu_{kl}(t)$\,. Therefore, when growing a network by connecting the new node to two existing ones, one should only make sure that the nodes are selected uniformly at random to achieve the best scaling. Of course, the latter is true as long as the approxiamtion in Eq.~(\ref{eqrtpu}) holds. If instead one wants to disrupt the system, a scaling of the Kirchhoff index $N_t$ times worth compared to the previous situation is obtained by maximizing either $\rho_{kl}(t) - \mu_{kl}(t)$ or $\rho_{kl}(t)$\,. The latter can be achieved by choosing nodes that are close in terms of $\Omega_{kl}(t)$ and $\Omega_{kl}^{(2)}(t)$ and rather peripheral in the network i.e. small $C(k,t)$\,, $C(l,t)$\,.

\subsection*{Remark}
One has to be careful when interpreting Eqs.~(\ref{eqm2}), (\ref{eqm13}), and notice that the Kirchhoff index increases at least linearly with $N_t$ on average. This can be seen from Eq.~(\ref{ineq}). More intuitively, in the case $m=N_t$\,, which means that the number of edges added at each iteration grows with the system size, assuming an initial all-to-all network one has,
\begin{eqnarray}
Kf_1{(t+1)} = {N_t}\,,
\end{eqnarray}
which monotonically increases. In this situation, one adds as many new paths between the existing nodes as possible when including a single new node. Therefore, for any $m<N_t$\,, the Kirchhoff index must increase. In Eq.~(\ref{eqm2}), one might however reduce the amplitude of the increase or sometimes even decrease $Kf_1$ by carefully selecting $k$ and $l$\,. But the latter can only be true for a few iterations.


\section{Conclusion}\label{sec4}
We considered the evolution of random networks where at each iteration, a new node is added and connected to one or two existing ones. When the new nodes only connects to one existing node, the scaling of $Kf_1(t)/N_t$ with the number of nodes is between $N_t$ and $N_t^2$ as the number of iterations becomes large. When the existing node is randomly uniformly chosen, the scaling is only logarithmically worse than the lower bound, i.e. $Kf_1(t)/N_t\overset{t\rightarrow\infty}{\propto}N_t\,\log N_t$\,. In the more complex situation where the new nodes connects to two existing ones, we derived a recursive expression for the evolution of $Kf_1(t)$\,. The latter is essentially given by $\rho_{kl}(t) - \mu_{kl}(t)$ which can be expressed in terms of the resistance distances and centralities i.e. $\Omega_{kl}$\,, $\Omega_{kl}^{(2)}$\,, $C^{-1}(k,t)$\,, $C^{-1}(l,t)$\,, [see Eqs.(\ref{eqmu}), (\ref{eqrho})]. We showed that, by introducing a bias in the selection of $k$ and $l$ toward the minimum/maximum of these quantities, one can tune the increase of $Kf_1(t)/N_t$ from linear to quadratic in $N_t$\,. 
For $m>2$\,, it is much more challenging to obtain analytical expression for the evolution of $Kf_1$\,. The same applies to the case where $m$ is a function of the number of nodes. Using the lower bound in Eq.~(\ref{ineq}) yet allows one to obtain the minimal scaling of the Kirchhoff index by correctly choosing $n_e(N)$\,.

The scenario we consider here applies to evolving systems where a single new node is added at each iteration and connects to existing nodes. This would represent the case where a new molecule forms bonds with another group of molecules, or an autonomous vehicle that joins a platoon by interacting with one or many of its members. With the results presented here, one can anticipate the scaling of the Kirchhoff index based on how the new units connect to the existing ones. Thus, they also give insights on the evolution  of fluctuations in networked systems such as consensus dynamics and synchronized systems that are diffusively coupled. 
\subsection{Outlook}
In this manuscript, we considered fundamental mechanisms for growing a networked systems, namely (i) adding edges to an existing system; (ii) adding nodes that connect to one or two existing units in the system. We investigated the two scenarios (i) and (ii) independently and yet found that different scalings for the Kirchhoff index are achievable. In order to describe realistic systems such as swarm formation in groups of animals or autonomous robots, one would have to consider the two mechanisms (i), (ii) occurring potentially one after the other or even simultaneously.
Future research should also consider the extension of these results to the case where multiple connected nodes are added at the same time. Additionally, one should investigate how other properties are modified by the growth of the network. Indeed, the Kirchhoff index is directly related to the small fluctuations of networked oscillators, but other system characteristics such as the ability of a network to synchronize typically depend on the maximum and minimum eigenvalues of the Laplacian matrix.

\section*{Ackowledgements}
This work has been supported by the Laboratory Directed Research and Development program of Los Alamos National Laboratory under project numbers 20220797PRD2 and 20220774ER and by U.S. DOE/OE as part of
the DOE Advanced Sensor and Data Analytics Program.




%

\end{document}